\documentclass[aps, prb, amsmath,amssymb, reprint, showpacs,superscriptaddress,groupedaddress, nofootinbib ]{revtex4-1}

\usepackage{nicefrac}
\usepackage{float}
\usepackage{graphicx}
\usepackage{dcolumn}
\usepackage{bm}
\usepackage{comment,caption,subcaption}
\usepackage{soul}
\usepackage{hyperref}
\hypersetup{
	colorlinks=true,
	citecolor=blue,
	linkcolor=blue,
	filecolor=blue,      
	urlcolor=blue,
}

\begin{document}
	
\title{Spin Gapless Semiconducting Nature in Co-rich $\mathrm{Co_{1+x}Fe_{1-x}CrGa}$: Insight and Advancements}

\author{Deepika Rani$^{1}$, Enamullah$^{1}$, Lakhan Bainsla$^{1,2}$, K. G. Suresh$^1$ and Aftab Alam$^{1}$}

\email{aftab@iitb.ac.in}

\affiliation{$^1$Department of Physics, Indian Institute of Technology Bombay, Powai, Mumbai 400076, Maharashtra, India \\ $^2$WPI Advanced Institute for Materials Research, Tohoku University, Sendai 980-8577, Japan }

\date{\today}

\begin{abstract}
	Spin gapless semiconductors (SGS) are interesting class of materials which bridge the gap between semiconductors and half-metallic ferromagnets. This class of materials shows band gap in one of the spin channels and a zero band gap in the other, and thus promote tunable spin transport. Here, we present structural, electronic, magnetic and transport properties of Co-rich spin gapless semiconductor CoFeCrGa using both theoretical and experimental techniques. The key advantage of Co-rich samples $\mathrm{Co_{1+x}Fe_{1-x}CrGa}$ is the high Curie temperature (T$\mathrm{_C}$) and magnetization, without compromising the SGS nature (up to x = 0.4), and hence our choice. The quaternary Heusler alloys $\mathrm{Co_{1+x}Fe_{1-x}CrGa}$ (x = 0.1 to 0.5) are found to crystallize in LiMgPdSn-type structure having space group $F\bar{4}3m$ (\# 216). The measured Curie temperature increases from 690 K (x = 0) to 870 K (x = 0.5). The obtained $\mathrm{T_C}$ for x = 0.3 (790 K) is found to be the highest among all the previously reported SGS materials. Observed magnetization values follow the Slater-Pauling rule. Measured electrical resistivity, in the temperature range of 5-350 K, suggests that the alloys retain the SGS behavior up to x = 0.4, beyond which it reflects metallic character. Unlike conventional semiconductors, the conductivity value ($\mathrm{\sigma_{xx}}$) at 300 K lies in the range of 2289 S $\mathrm{cm^{-1}}$ to 3294 S $\mathrm{cm^{-1}}$, which is close to that of other reported SGS materials. The anomalous Hall effect is comparatively low. The intrinsic contribution to the anomalous Hall conductivity increase with x, which can be correlated with the enhancement in chemical order. The anomalous Hall coefficient is found to increase from 38 S/cm for x = 0.1 to 43 S/cm for 0.3. Seebeck coefficients turn out to be vanishingly small below 300 K, another signature for being SGS. All the alloys (for different x) are found to be both chemically and thermally stable. Simulated magnetization agrees fairly with the experiment. As such Co-rich CoFeCrGa is a promising candidate for room temperature spintronic applications, with enhanced T$\mathrm{_C}$, magnetic properties and SGS nature.
	
	\end{abstract}
\pacs{85.75.−d, 72.20.−i, 75.50 Pp, 76.80.+y, 72.15.-v, 71.15.Mb, 71.20.-b}
\keywords{Heusler alloys, Density Functional Theory, Transport properties, Magnetism, Spin Polarization, Spin-gapless semiconductors, Half-metals}
\maketitle

\section{Introduction} 
\begin{figure}
\centering
\includegraphics[width=0.9\linewidth]{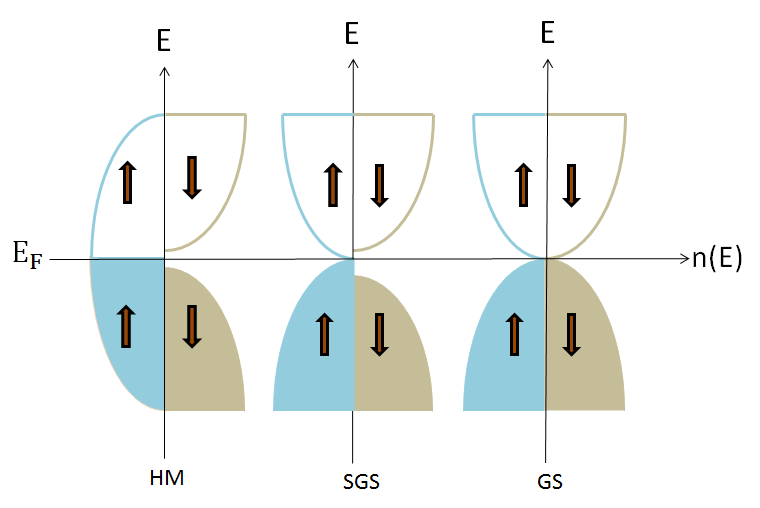}
\caption{Schematic representation of density of states of (a) Half-metal (b) Spin-gapless Semiconductor and (c) Gapless Semiconductor.}
\label{dos}
\end{figure}

In recent years, a new class of materials known as spin gapless semiconductors (SGS) has attracted a  lot of attention due to their peculiar electronic structure and potential applications in spintronic devices. SGS materials exhibit a finite band gap for one spin channel and a closed gap for the other spin. \cite{GS} These materials can be regarded as a combination of half-metallic ferromagnets and gapless semiconductors. This feature results in peculiar transport properties and applications in the field of spintronics since the conducting carriers (electrons or holes) are not only completely spin polarized but also easily excited due to the gapless nature in one of the spin-bands. Also, mobility of carriers in this class of materials is considerably higher than that of conventional semiconductors. The schematic representation of a half-metal, gapless semiconductor and spin-gapless semiconductor is shown in Fig.\ref{dos}. SGS behavior was predicted in dilute magnetic semiconductor(DMS) PbPdO$_2$ by fisrt principles calculation. \cite{WANG201555} However, one of the major drawbacks of DMS is the low Curie temperature ($\mathrm{T_C}$). \cite{doi:10.1063/1.2992200} Heusler alloy based SGS systems have advantages over DMS and other reported magnetic semiconductors. They have stable structure, high $\mathrm{T_C}$ and high spin-polarization, which make them suitable for applications in the field of spintronics. SGS behavior has been identified theoretically in many Heusler alloys \cite{0295-5075-102-1-17007, doi:10.1063/1.4775599, doi:10.1063/1.4871403, PhysRevLett.100.156404} but only a few have been confirmed experimentally. \cite{PhysRevB.92.045201, PhysRevB.91.104408, PhysRevLett.110.100401, PhysRevB.97.054407}

SGS behavior was verified in the equiatomic Heusler alloy CoFeCrGa(CFCG). \cite{PhysRevB.92.045201} This alloy was found to crystallize in cubic Heusler structure (LiMgPdSn prototype) with $\mathrm{DO_3}$ disorder with lattice parameter of 5.79 $\mathrm{\AA}$. The saturation magnetization was found to be 2.1 $\mu_B$/f.u. at 8 K. One of the methods to design new materials in the Heusler alloys $\mathrm{XX^{\prime}YZ}$ is by exchanging the elements X, X$^{\prime}$, Y and Z or by substituting them by other elements. We have substituted Co for Fe in CFCG with the objective of improving its properties such as the band gap, Curie temperature and spin-polarization, while trying to retain the SGS nature. This is because stable materials with large band gap, high spin polarization and high Curie temperature are desired for spintronic applications. We have synthesized $\mathrm{Co_{1+x}Fe_{1-x}CrGa}$ (x = 0.1 to 0.5) alloys and studied their structural, electronic, magnetic and transport properties. We found that all the alloys crystallize in cubic Heusler structure and the magnetization increases with excess Co and are in close agreement with the Slater-Pauling rule, a prerequisite for spintronic materials. Also, the Curie temperature of all the alloys is well above the room temperature. The alloys with x = 0.1 to 0.4 retain SGS nature whereas, it becomes half-metallic (HM) at x = 0.5, as indicated by the resistivity data. The extrinsic and intrinsic contributions are separated in Hall resistivity. The intrinsic contribution is found to increase with x and is correlated with the improved chemical order within the lattice. The extrinsic contribution is found to be negative and thus contributes to the anomalous Hall effect (AHE) in the opposite way as the Karplus-Luttinger term (Intrinsic contribution) contributes. The conductivity value ($\mathrm{\sigma_{xx}}$) at 300 K  lies in the range of 2289 S $\mathrm{cm^{-1}}$ to 3294 S $\mathrm{cm^{-1}}$, which is close to that of other reported SGS materials. The negligible Seebeck coefficient along with the conductivity behavior supports the SGS nature. Thus, the spin-gapless semiconducting nature of CoFeCrGa alloy is robust against the substitution of Co by Fe up to x = 0.4.

\section{Experimental Details}

\subsection{Sample Synthesis}
The polycrystalline alloys $\mathrm{Co_{1+x}Fe_{1-x}CrGa}$ were synthesized by arc melting the stoichiometric amounts of constituent elements (with at least 99.9\% purity) in water cooled copper hearth under high purity argon atmosphere. To further reduce the contamination, a Ti ingot was used as the oxygen getter. To ensure homogeneity, the ingot formed was  melted several times after flipping. After melting, the samples were sealed in a quartz tube and annealed for 7 days at 1073 K followed by furnace cooling. 

\subsection{Characterization}
To determine the crystal structure of the alloys X-ray diffraction (XRD) pattern was taken at room temperature using X”pert pro diffractometer with $\mathrm{Cu-K\alpha}$ radiation. XRD analysis was done with the help of FullProf suite \cite{RR} which exploits the least square refinement between the experimental and calculated intensities. Magnetization isotherms at 5 K were obtained using a vibrating sample magnetometer (VSM) attached to the physical property measurement system (PPMS) (Quantum design) for fields up to 50 kOe. Thermo-magnetic curves in the high temperature region (300 K -950 K) were obtained using a VSM attached with high temperature oven, under a field of 500 Oe. Resistivity measurements were carried out using a physical property measurement system (PPMS; Quantum Design) using the linear four probe method, applying a current of 5 mA. Hall measurement was performed using the PPMS with a five probe method by applying a current of 150 mA. 

\section{Computational details}
Electronic structure of the alloy, Co$_{1+x}$Fe$_{1-x}$CrGa for x=0.0,0.125,0.25,0.375 and 0.50 has been simulated using Density Functional Theory (DFT) \cite{PhysRev.136.B864}implemented in Vienna {\it{ab-initio}} Simulation Package (VASP) {\cite{PhysRevB.54.11169,KRESSE199615,PhysRevB.47.558}} with a
projected augmented wave(PAW) basis. \cite{PhysRevB.59.1758} In order to achieve the concentration 'x' close to the experiment, we made a symmetric 2$\times$2$\times$2 supercell of the primitive cell.
The generalized gradient approximations(GGA) along with the Perdew-Burke-Ernzerhof (PBE) \cite{PhysRevLett.77.3865} is adopted for the electronic exchange and correlation functional. A plane-wave energy cutoff of 500 eV is used with the energy convergence of 10$^{-6}$ eV/cell. For Brillouin zone integration, 12$\times$12$\times$12 k-points and accurate precession along with the conjugate gradient(CG) algorithm are used.

The lattice dynamics of the alloys has been analyzed using phonon frequencies calculated within the frame work of Density Functional Perturbation Theory(DFPT). \cite{RevModPhys.73.515,PhysRevB.55.10355} We generate the displaced atomic supercells using the open source PHONOPY software package\cite{PhysRevB.78.134106} and fit the force constant data to compute the dynamical matrix.

\section{Results and Discussion}

\begin{figure}
	\centering
	\includegraphics[width=0.9\linewidth]{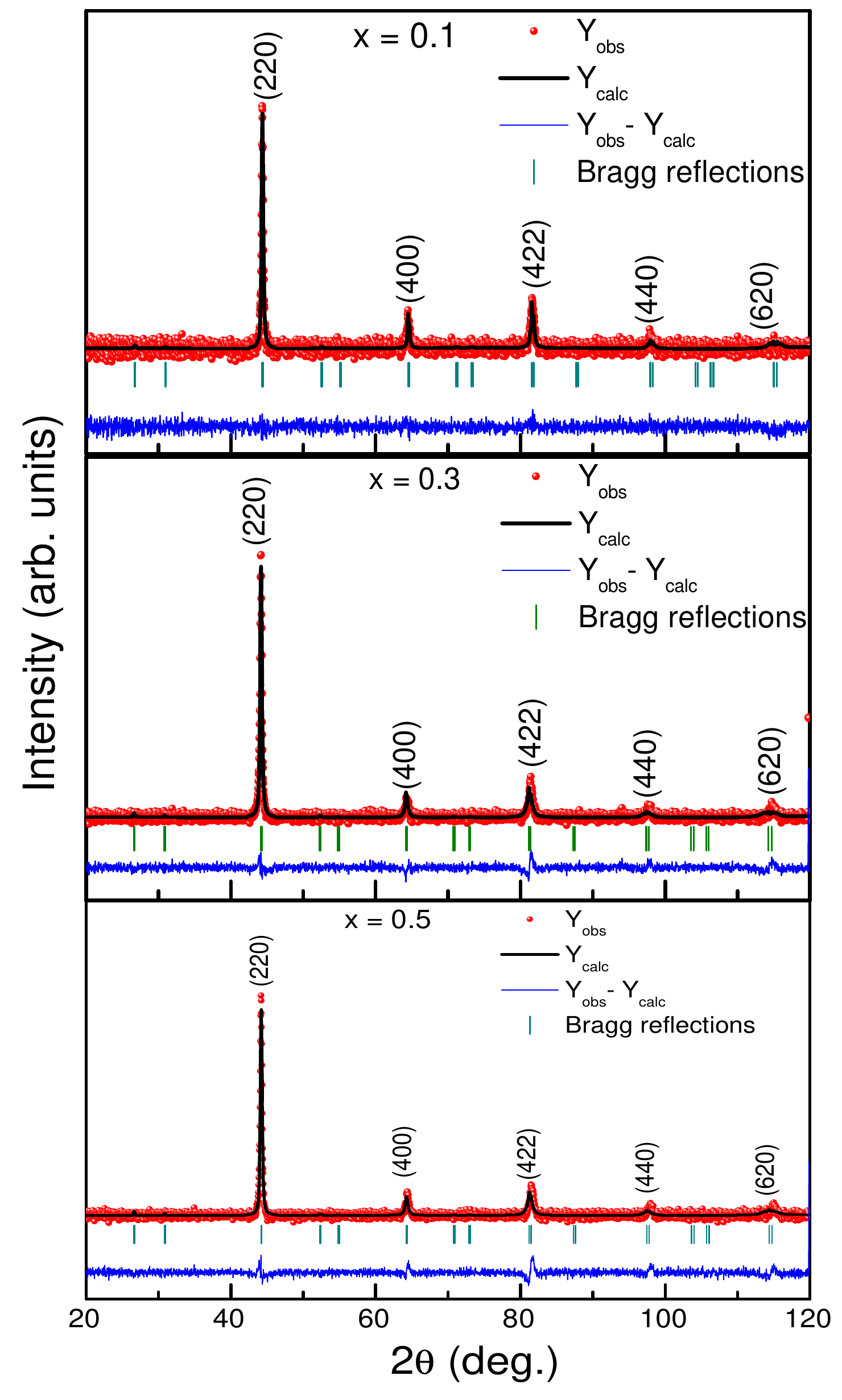}
	\caption{Rietveld refined XRD pattern of $\mathrm{Co_{1+x}Fe_{1-x}CrGa}$ (x = 0.1, 0.3 and 0.5) alloys. $\mathrm{Y_{obs}}$ and $\mathrm{Y_{calc}}$ are the observed and calculated scattered intensities. }
	\label{XRD}
\end{figure}

\begin{figure}
	\centering
	\includegraphics[width=0.9\linewidth]{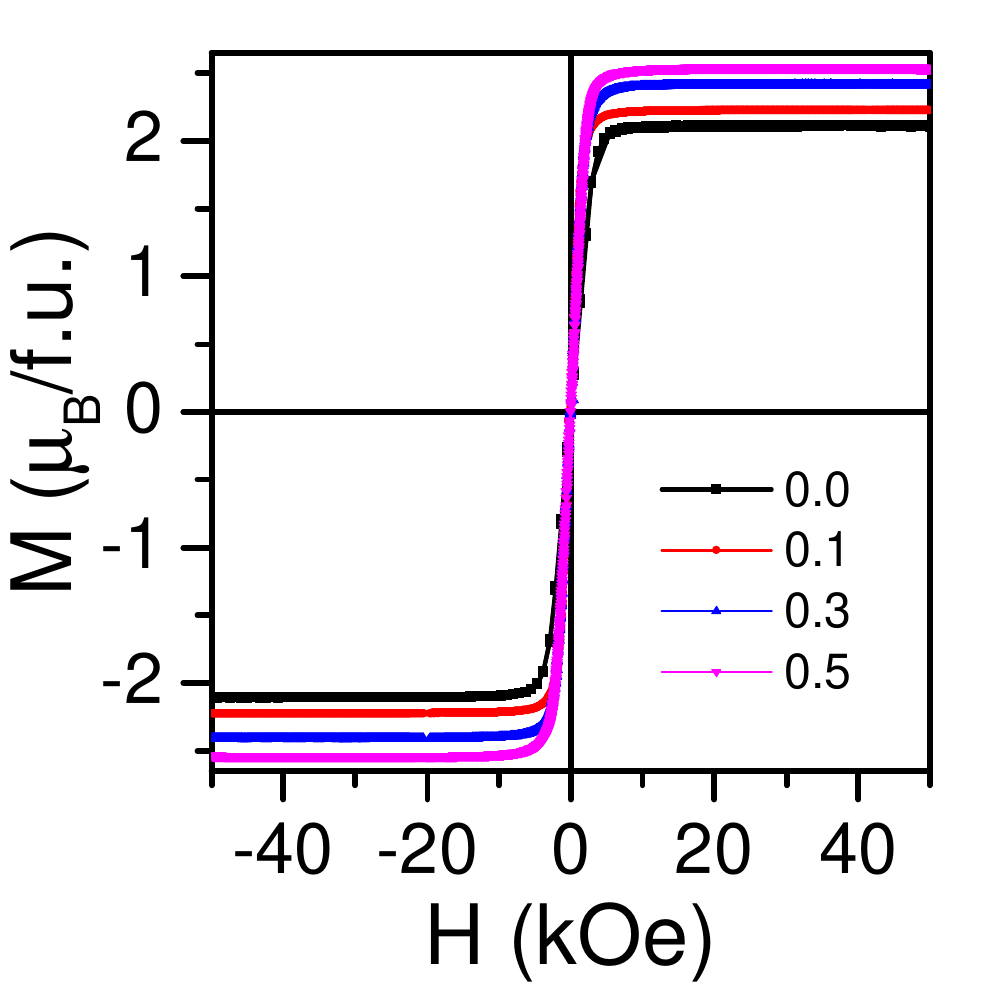}
	\caption{Isothermal magnetization curves for $\mathrm{Co_{1+x}Fe_{1-x}CrGa}$ alloys at 5 K.}
	\label{MH}
\end{figure}

\begin{figure}
	\centering
	\includegraphics[width=0.9\linewidth]{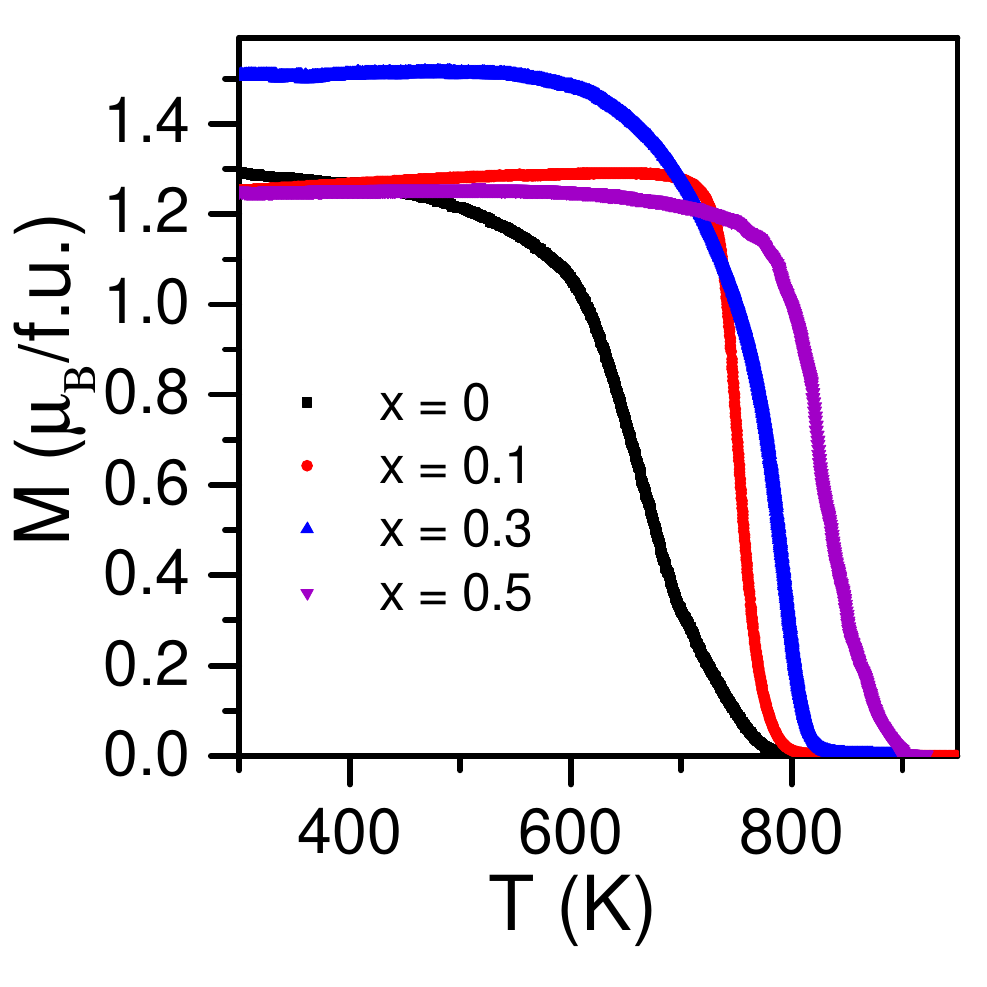}
	\caption{Temperature dependence of magnetization in 500 Oe for Co$_{1+x}$Fe$_{1-x}$CrGa alloys. }
	\label{MT}
\end{figure}
\begin{figure}
	\centering
	\includegraphics[width=1.0\linewidth]{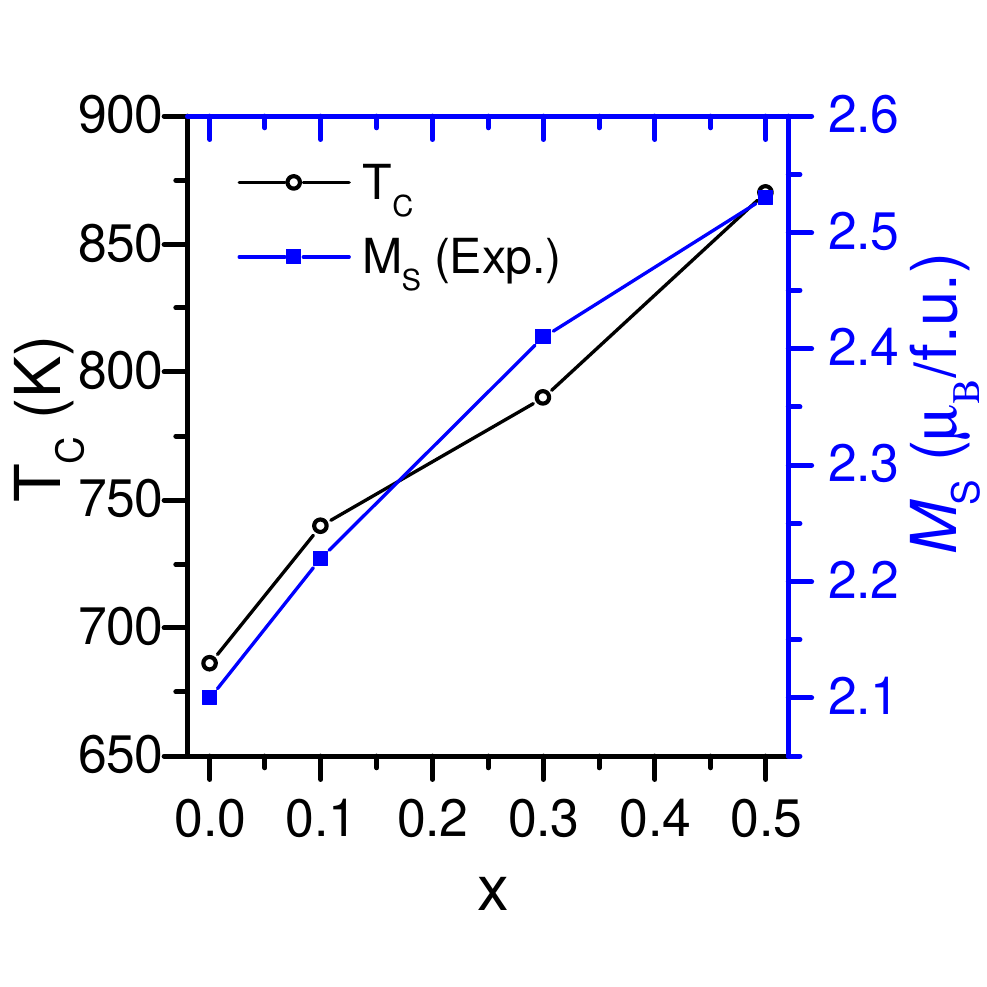}
	\caption{Variation of Curie temperature (T$_C$) and experimental saturation magnetization with x for Co$_{1+x}$Fe$_{1-x}$CrGa alloys. $T_c$ is calculated from the minima of the first order derivative of $M$ vs. $T$ curve.}
	\label{MT1}
\end{figure}

\begin{figure*}
	\centering
	\includegraphics[width=0.9\linewidth]{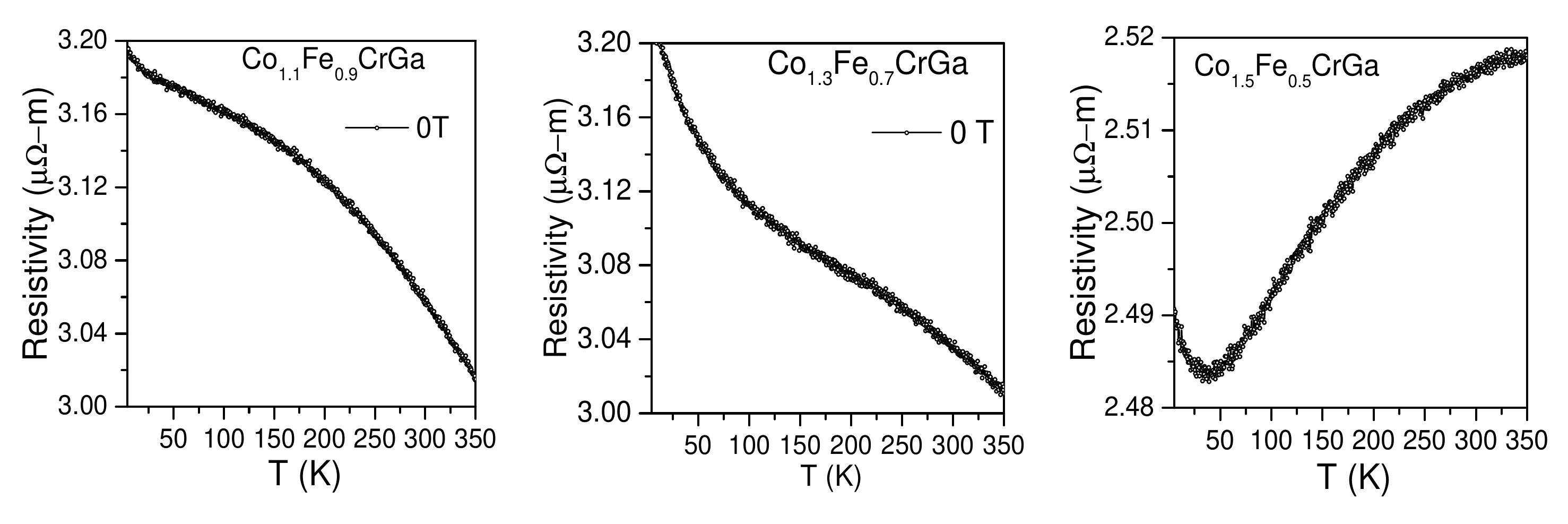}
	\caption{Temperature dependence of electrical resistivity($\rho$) for $\mathrm{Co_{1+x}Fe_{1-x}CrGa}$ alloys in zero field.}
	\label{Res}
\end{figure*}

\begin{figure}
	\centering
	\includegraphics[width=0.9\linewidth]{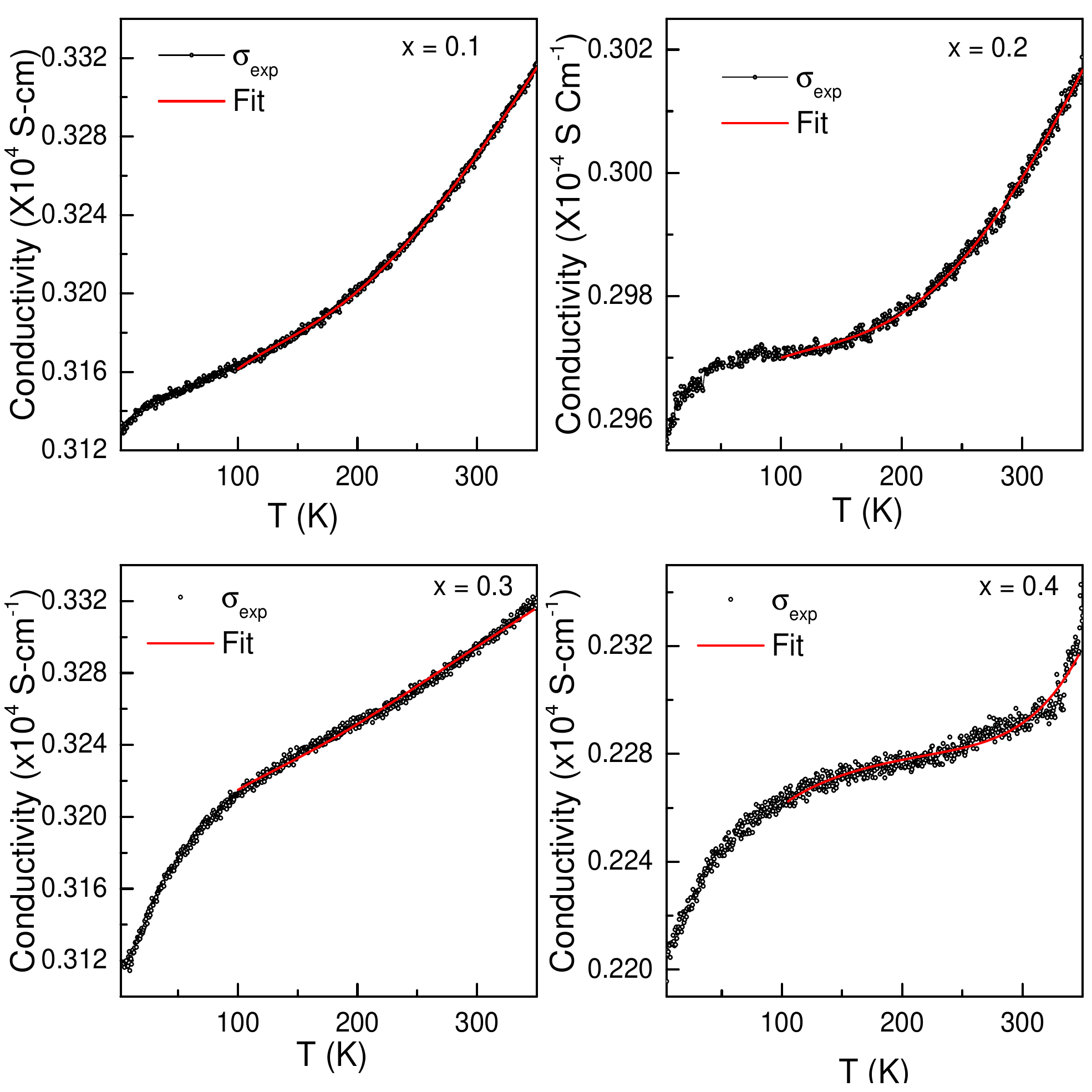}
	\caption{Variation of electrical conductivity with temperature in zero field along with the fit (using Eq. (6)) shown by red line (above 100 K) using two carrier model for $\mathrm{Co_{1+x}Fe_{1-x}CrGa}$  (x = 0.1, 0.2, 0.3 and 0.4).}
	\label{cond}
\end{figure}

\begin{figure*}
\centering
\includegraphics[width=0.9\linewidth]{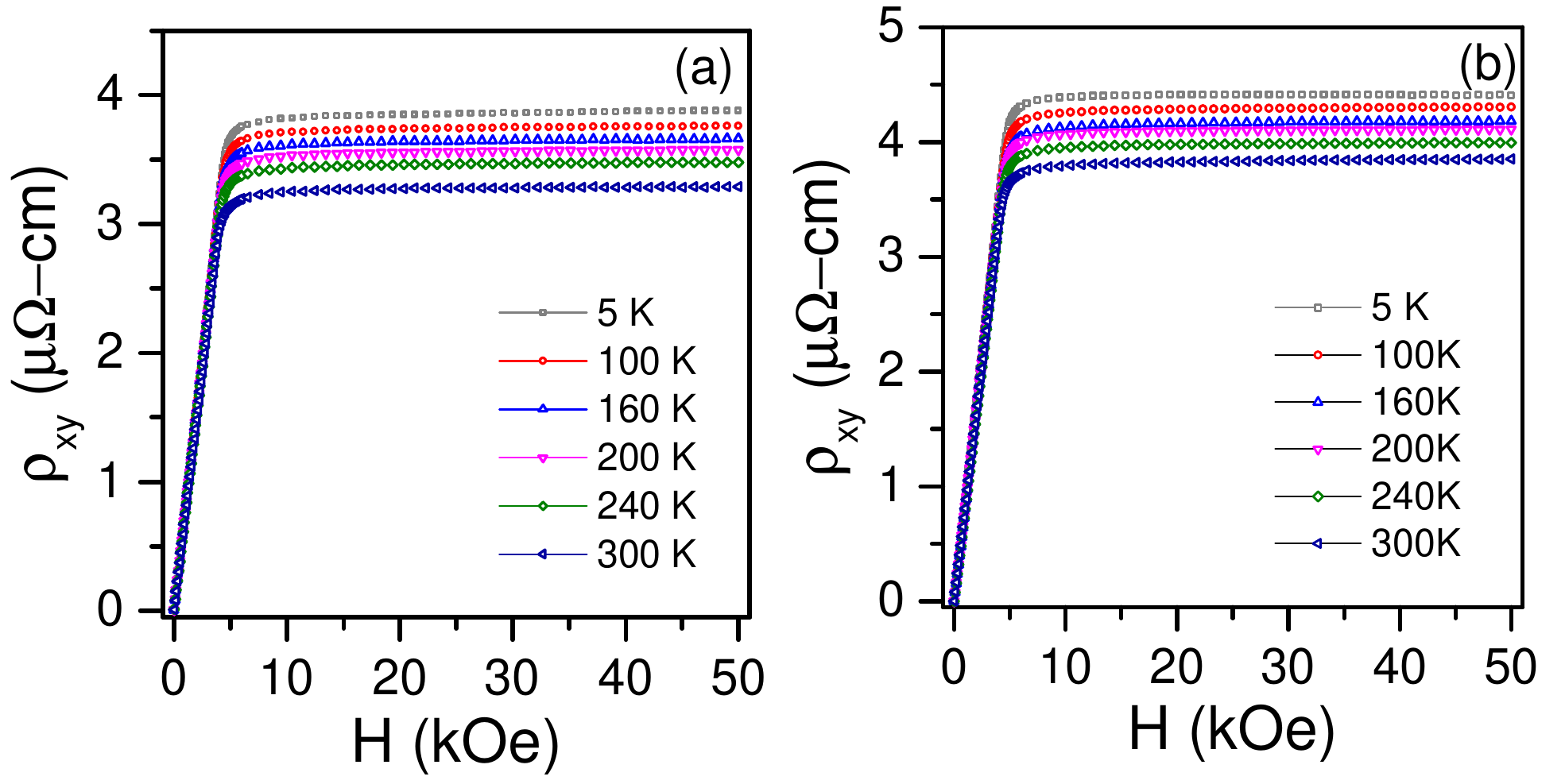}
\caption{Hall resistivity ($\rho_{xy}$) as a function of magnetic field(H), measured at different temperatures for (a) $\mathrm{Co_{1.1}Fe_{0.9}CrGa}$ and (b)$\mathrm{Co_{1.3}Fe_{0.7}CrGa}$.}
\label{hall_temp}
\end{figure*}

\begin{figure}
\centering
\includegraphics[width=0.9\linewidth]{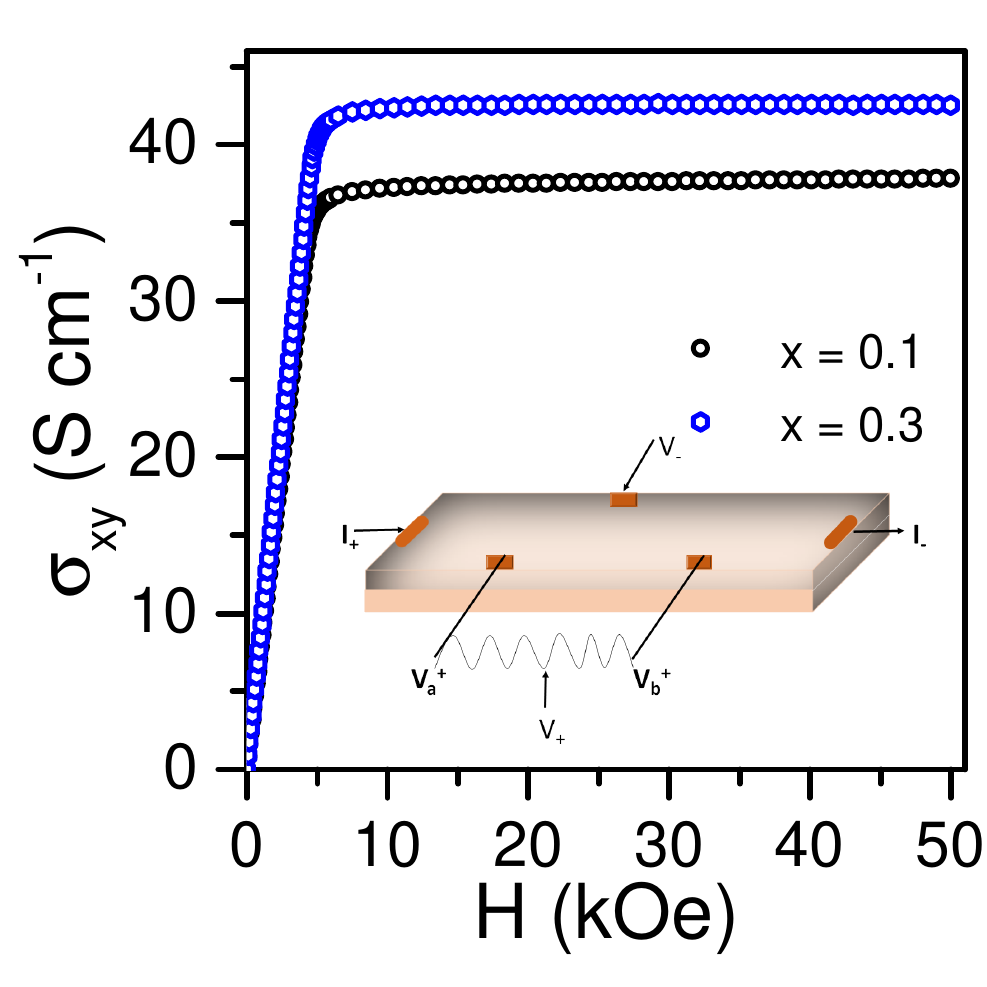}
\caption{Hall conductivity $\mathrm{\sigma_{xy}}$ versus applied field for $\mathrm{Co_{1+x}Fe_{1-x}CrGa}$ (x = 0.1 and 0.3) at 5 K.}
\label{fig:sigma_all}
\end{figure}

\begin{figure*}
	\centering
	\includegraphics[width=0.9\linewidth]{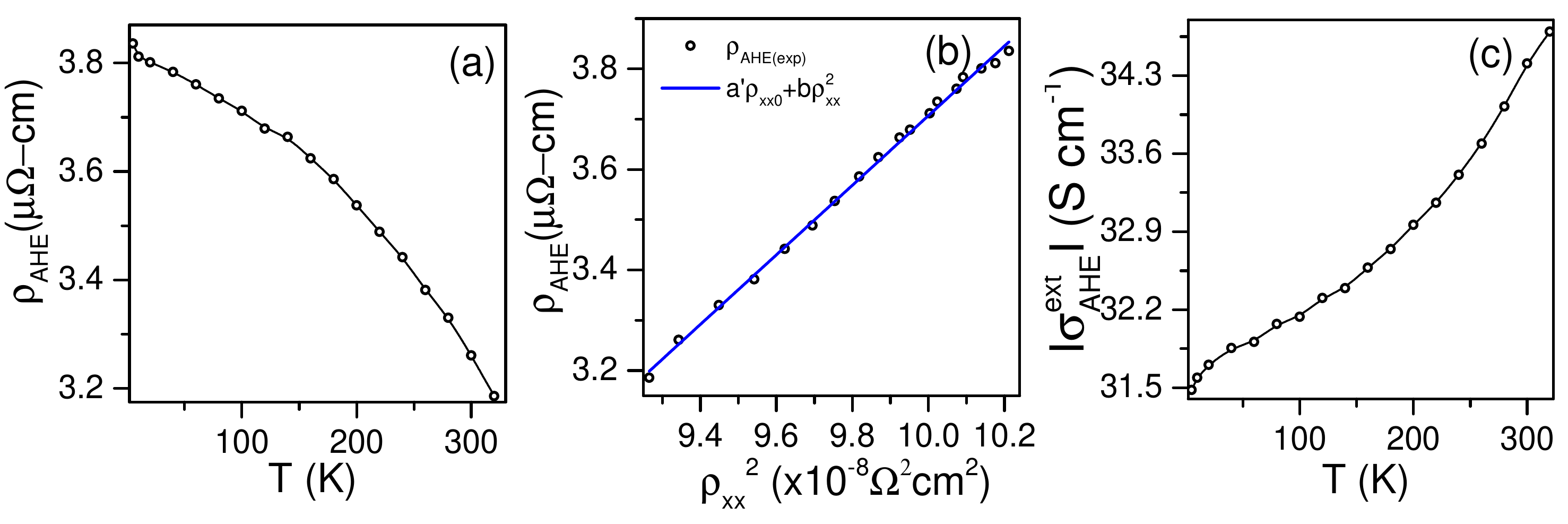}
	\caption{The variation of (a) $\rho_{AHE}$ with temperature (b) $\rho_{AHE}$ with $\rho_{xx}^2$ with the fitted curve (blue) using Eq. (9) and (c) extrinsic contribution of $\sigma_{AHE}$ with temperature for $\mathrm{Co_{1.1}Fe_{0.9}CrGa}$ alloy }
	\label{fig:CFCG1_hall}
\end{figure*}

\subsection{Experimental Results}

\subsubsection{X-ray diffraction}
\label{XRayD}
Figure \ref{XRD}  shows the Rietveld refined room temperature XRD patterns of $\mathrm{Co_{1+x}Fe_{1-x}CrGa}$ alloys (x = 0.1, 0.3 and 0.5). Similar patterns were obtained for x = 0.2 and 0.4 (not shown here). It is clear from the pattern that the alloys crystallise in the cubic structure. The quaternary Heusler alloys, XX$^{\prime}$YZ exhibit LiMgPdSn-type structure whose primitive cell contains four atoms at the Wyckoff positions 4a, 4b, 4c and 4d. The structure factor for the quaternary Heusler alloy with X at 4b, X' at 4c, Y at 4d and Z at 4a Wyckoff positions can be written as:\\
\begin{equation}
\small
F_{hkl}=4(f_Z+f_Y e^{\pi i(h+k+l)}+f_X e^{\frac{\pi}{2} i(h+k+l)}+f_{X'} e^{\frac{-\pi}{2} i(h+k+l)})
\end{equation}
where, ${f_X, f_{X'}, f_Y}$ and ${f_Z}$ are the atomic scattering factors for X, X', Y and Z respectively.
The structure factor for the superlattice reflections can be written as:
\begin{equation}
\mathrm{F_{111}} = 4[\mathrm{(f_Y-f_Z)-i(f_X-f_{X'})}]
\end{equation}
\begin{equation}
\mathrm{F_{200}} = 4[\mathrm{(f_Y+f_Z)-(f_X-f_{X'})}]
\end{equation}
In the case of B2 disorder(Y and Z atoms are randomly distributed), the intensity of the (111) peak should reduce or disappear as seen from equation (2). For a completely disordered structure i.e., A2-type (all the four atoms occupy random positions), both the superlattice peaks (111) and (200) should be absent.
It is found that the Co substitution in place of Fe has not changed the crystal structure of the parent compound CoFeCrGa. The low angle, order dependent superlattice reflections (111) and (200) peaks, which are characteristics of a perfectly ordered Heusler structure were not visible in the XRD pattern. This could be due to anti-site disorder (B2-type, $\mathrm{DO_3}$ or A2-type) or the similar scattering factors of the constituent elements Co, Fe, Cr and Ga.
The earlier report on M\"{o}ssbauer spectroscopy studies of CoFeCrGa alloy has confirmed the $\mathrm{DO_3}$ disorder.\cite{PhysRevB.92.045201} Thus, the alloys in the present investigation may have some amount of $\mathrm{DO_3}$ disorder, though a complete information of order is not possible using XRD alone for these alloys. The lattice parameters were found using Fullprof Suite software assuming the Y - type ordering. The lattice parameters were found to be 5.79, 5.78 and 5.76 $\mathrm{\AA}$ for x = 0.1, 0.3 and 0.5 respectively. Thus, substituting Fe by Co leads to only a marginal change in the lattice parameter.

\begin{figure*}
	\centering
	\includegraphics[width=0.9\linewidth]{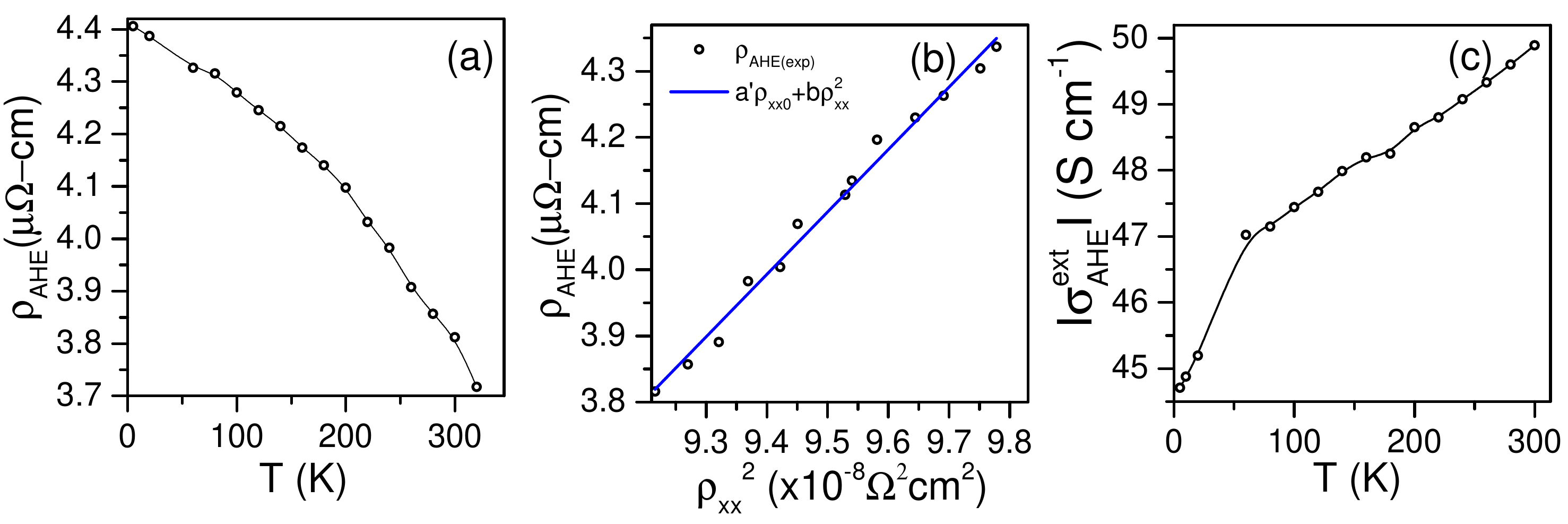}
	\caption{The variation of (a) $\rho_{AHE}$ with temperature (b) $\rho_{AHE}$ with $\rho_{xx}^2$ with the fitted curve (blue) using Eq. (9) and (c) extrinsic contribution $\sigma_{AHE}$ with temperature for $\mathrm{Co_{1.3}Fe_{0.7}CrGa}$ alloy.}
	\label{fig:CFCG3_hall}
\end{figure*}

\begin{figure*}
	\centering
	\includegraphics[width=0.9\linewidth]{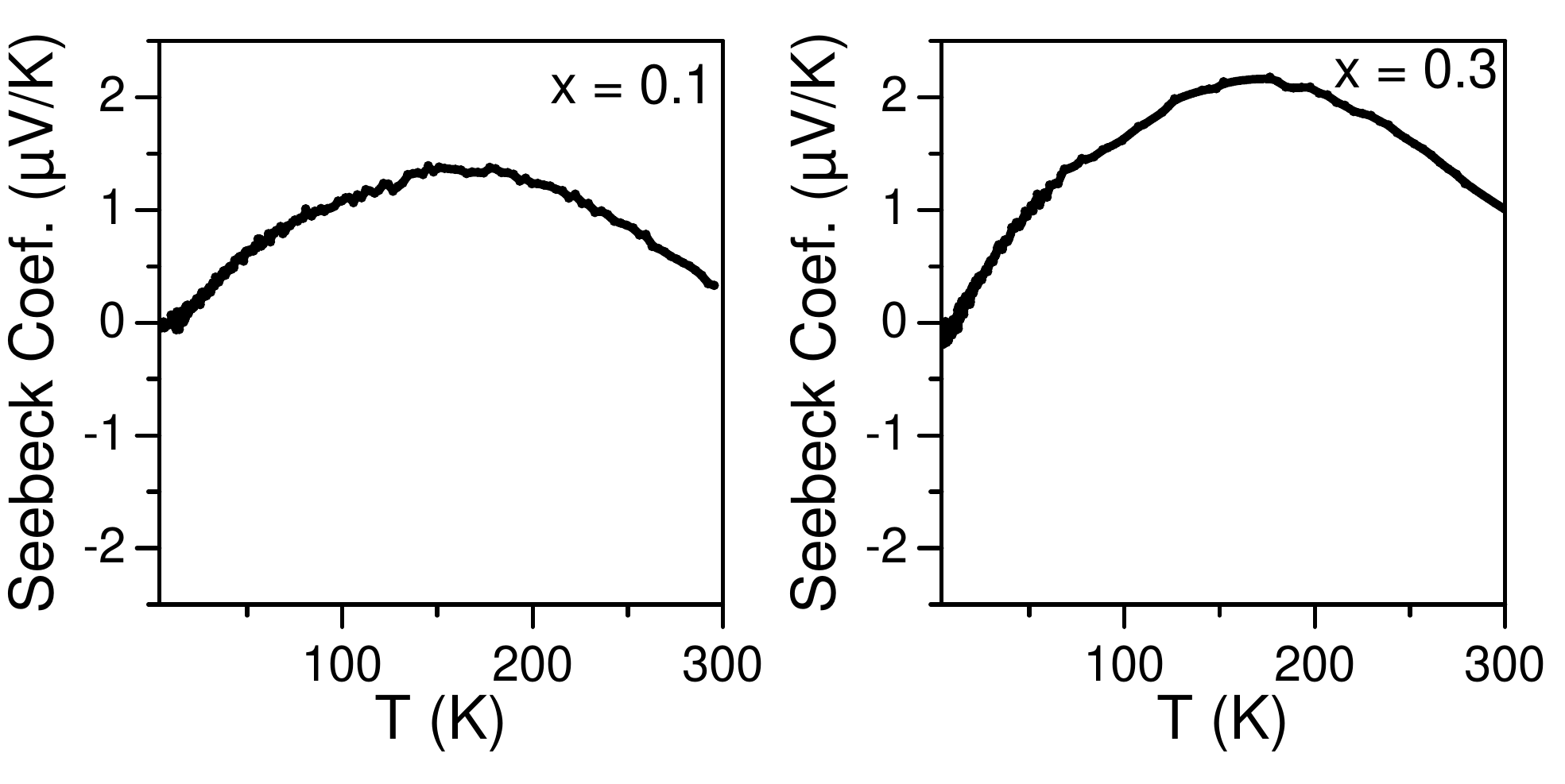}
	\caption{Variation of Seebeck coefficient with temperature for $\mathrm{Co_{1.1}Fe_{0.9}CrGa}$ and $\mathrm{Co_{1.3}Fe_{0.7}CrGa}$ alloy.}
	\label{SC}
\end{figure*}

\begin{figure*}
	\centering
	\includegraphics[width=0.9\linewidth]{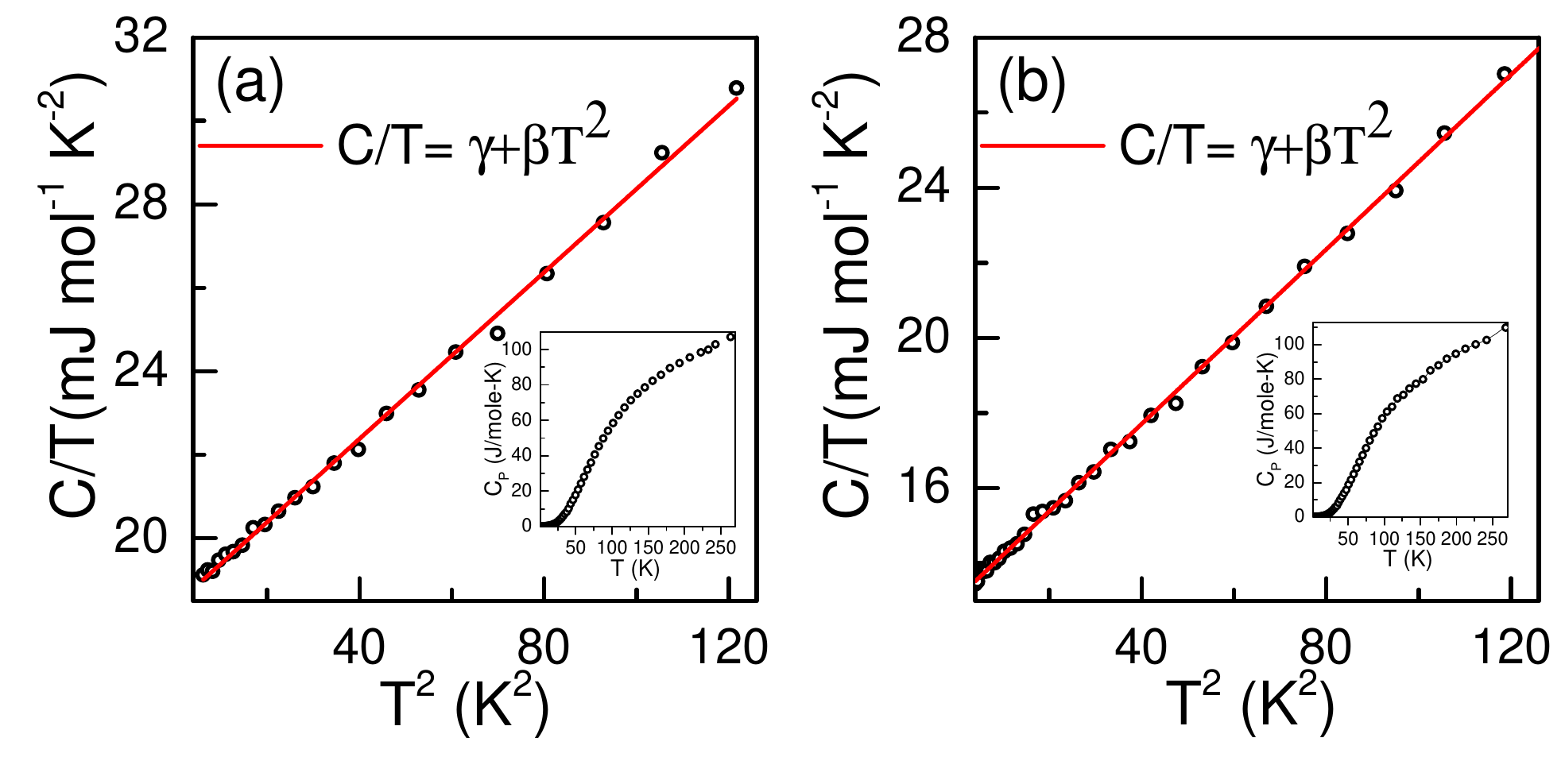}
	\caption{Heat capacity (C/T) vs.$\mathrm{T^2}$ plot for (a)$\mathrm{Co_{1.1}Fe_{0.9}CrGa}$ and (b)$\mathrm{Co_{1.3}Fe_{0.7}CrGa}$, in low temperature range. Insets shows the C vs. T curve for high T range, up to 250 K.}
	\label{HC}
\end{figure*}

\begin{figure*}
	\centering
	\includegraphics[width=0.9\linewidth]{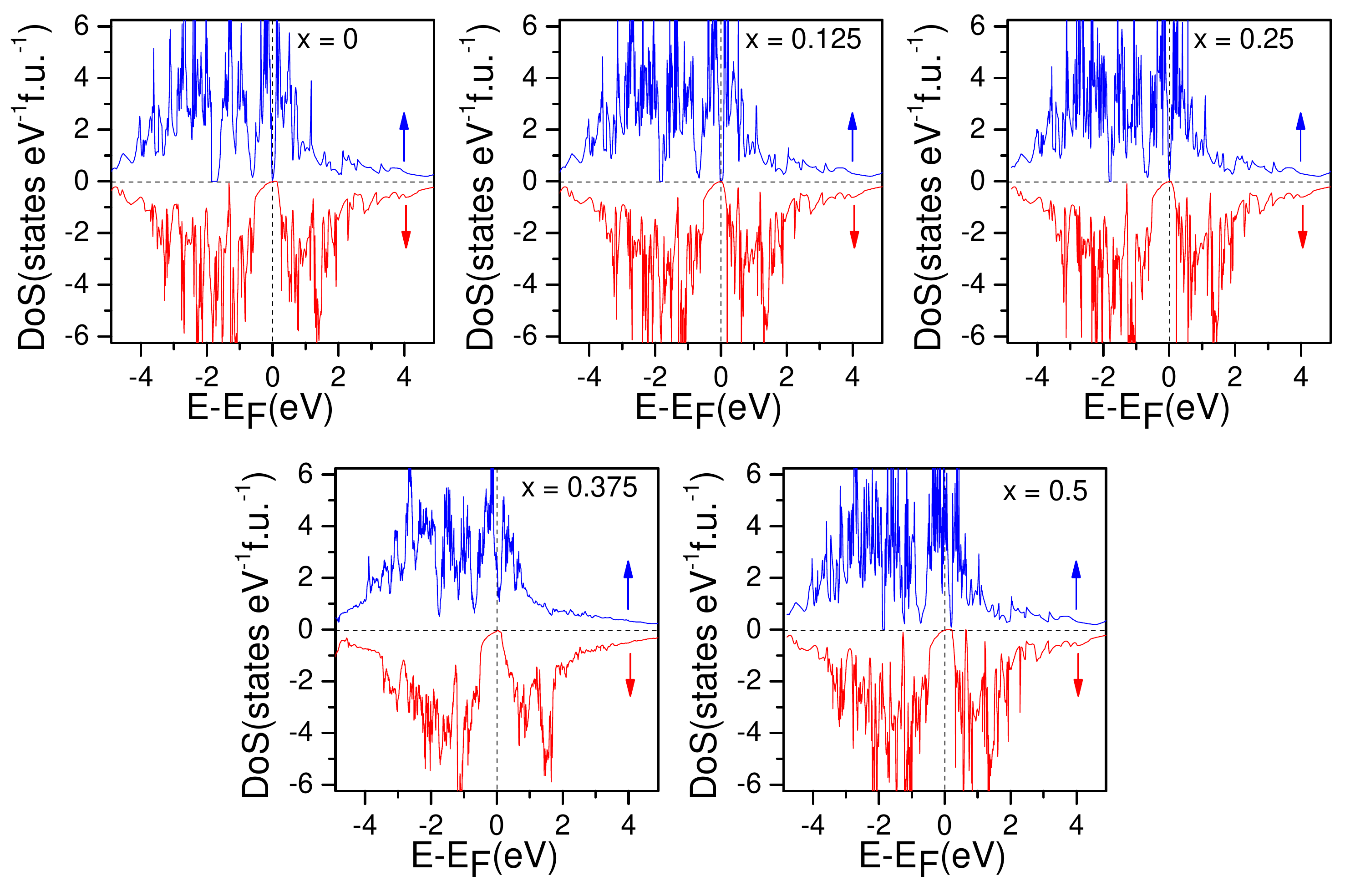}
	\caption{Spin polarized density of states for Co$_{1+x}$Fe$_{1-x}$CrGa at different x. }
	\label{dosall}
\end{figure*}

\subsubsection{Magnetic Properties}
Figure \ref{MH} shows the isothermal magnetization curves for $\mathrm{Co_{1+x}Fe_{1-x}CrGa}$ (x = 0, 0.1, 0.3 and 0.5) alloys at 5 K. The almost negligible hysteresis shows the soft magnetic nature of the alloys. Heusler alloys are known to follow the Slater-Pauling rule,\cite{PhysRevB.66.174429, PhysRevB.96.184404} according to which the total moment per unit cell is given by:
\begin{equation}
m = (N_v - 24) \mu_B
\end{equation}
where $N_v$ is the number of valence electrons per unit cell.\\
Since the valance electronic configuration of Co, Fe, Cr and Ga are $\mathrm{4s^23d^7,4s^23d^6, 4s^13d^5 and 4s^24p^1}$ respectively,  as per Slater-Pauling rule, the magnetic moment should increase from 2.1$\mathrm{\mu_B/f.u.}$ for x = 0.1 to 2.5 $\mathrm{\mu_B/f.u.}$ for x = 0.5.  The obtained experimental $M_s$ values (at 5 K) are in fair agreement with those calculated using the Slater-Pauling rule.

Figure \ref{MT} shows the temperature dependence of the magnetization  at a constant field of 500 Oe in the temperature range of 300-950 K for $\mathrm{Co_{1+x}Fe_{1-x}CrGa}$ alloys. The Curie temperature was estimated by taking the minima of the first order derivative of M - T curve. Figure \ref{MT1} shows the Curie temperature vs. x curve, which shows that the T$_\mathrm{C}$ increases almost linearly with x. It is found to be 686 K for x = 0 and  870 K for x = 0.5. The obtained $\mathrm{T_C}$ value for x = 0.3 (790 K) is found to be the highest among all the previously reported SGS materials. \cite{PhysRevB.91.104408, PhysRevLett.110.100401, PhysRevB.97.054407} The high $\mathrm{T_C}$ makes these alloys better suited for spintronic applications in real-time devices.

 \subsubsection{Resistivity Measurements: Longitudinal component}

Figure \ref{Res} shows the measured temperature dependence of the longitudinal resistivity ($\rho_{xx}$) for $\mathrm{Co_{1+x}Fe_{1-x}CrGa}$ alloys in zero field. The resistivity behaviour in the temperature region from 5 - 350 K was investigated.
\begin{quote}
	\centering
I. \hspace{0.3cm}{x = 0.1, 0.2, 0.3 and 0.4}
\end{quote}
As shown in Fig.\ref{Res}(a) and \ref{Res}(b), alloys with x = 0.1 and 0.3 show semiconducting behavior with a negative temperature coefficient ($d\rho/dT < 0$). A similar dependence was observed for x = 0.2 and 0.4. The temperature dependence of electrical conductivity $\sigma_{xx}$ for $\mathrm{Co_{1+x}Fe_{1-x}CrGa}$ (x = 0.1, 0.2, 0.3 and 0.4) is shown in Fig.\ref{cond} and it is clear that the alloys with x = 0.1 to 0.4 exhibit non-metallic behavior; $\sigma_{xx}$ increases with increasing temperature. The $\sigma_{xx}$ vs. T behavior is different from that of conventional metals or semiconductors as they exhibit exponential decrease or increase in the conductivity. A similar kind of behavior is also reported for the parent alloy CoFeCrGa. \cite{PhysRevB.92.045201} Thus, the $\sigma_{xx}$ vs. T behavior gives an indication of SGS nature in these alloys. The value of $\sigma_{xx}(T)$ at 300 K lies in the range 2290 $\mathrm{S cm^{-1}}$ to 3294 $\mathrm{S cm^{-1}}$ and is close to that of other reported SGS materials. \cite{PhysRevB.92.045201, PhysRevB.91.104408} To further analyze this behavior, $\sigma_{xx}$(T) is described by two-carrier model\cite{doi:10.1063/1.4823601,PhysRevApplied.7.064036,PhysRevB.97.054407} in the temperature range T = 100 to 350 K. In this model, the total conductivity is given by:
\begin{equation}
\frac{1}{\rho_{xx}}=\sigma_{xx} (T)=e(n_e\mu_e+n_h\mu_h)
\end{equation}
where, the first and second terms correspond to the electronic and hole components respectively. The mobilities are given by $\mu_i=\frac{1}{\alpha_i T + \beta_i}$ where i = e or h. Here, the $\alpha$-term corresponds to electron phonon scattering and $\beta$ term results from the mobility due to defects at T = 0 K. The fit assumes that the number of carriers is given by $n_e \sim n_{e0}\exp(-\Delta E_e/k_BT)$ and $n_h \sim n_{ho}\exp(-\Delta E_h/k_BT)$, where $\Delta E_i$ are the gaps for the carriers. Thus, under these assumptions, the conductivity can now be written as:

\begin{equation}
\small
\sigma_{xx} (T)= A(T)\exp(-\Delta E_e/K_BT)+B(T)\exp(-\Delta E_h/K_BT)
\end{equation}

where, $\mathrm{A(T)=\frac{en_{e0}\mu_{e0}}{1+\alpha_e'T}}$ and $\mathrm{B(T)=\frac{en_{h0}\mu_{h0}}{1+\alpha_h'T}}$.\\
Fitting to this model (red solid curves in Fig.\ref{cond}) gives the temperature coefficients $\alpha_e' \sim 0$ and $\alpha_h' \sim 0 $, which suggest that in these alloys the mobility is mainly due to defect scattering instead of phonons. The gaps $\Delta E_e (\Delta E_h)$ are found to be 76.2 meV (0.1), 89.5 meV (0.1) and 53.5 meV (0.1) for x = 0.1, 0.2 and 0.3 respectively.
Here, the low gaps are associated with holes since from the Hall measurement (see Fig. \ref{hall_temp}), it is found that holes are the majority carriers.

\begin{quote}
	\centering
	II. \hspace{0.3cm}{x = 0.5}
	\end{quote}

For x = 0.5, the resistivity is found to increase with temperature and thus shows the metallic behavior(Fig. \ref{Res}(c)). A minimum in the resistivity with an upturn is observed in this alloy. Such behavior is observed in some other half-metallic Heusler alloys \cite{doi:10.1063/1.4862966, PhysRevApplied.10.054022} as well and is generally ascribed to a weak localization mechanism i.e., the disorder augmented coherent backscattering of conduction electrons. \cite{RevModPhys.57.287} Thus, transition from SGS to HM behavior is expected when 50\% Fe is replaced by Co in  CoFeCrGa.

\subsubsection{Anomalous Hall effect measurements}

For a ferromagnet, in addition to the ordinary Hall effect(OHE), the transverse resistivity has a contribution from the magnetization as well, known as anomalous Hall resistivity. This additional transverse voltage (known as anomalous Hall effect) is due to the asymmetric scattering of current carrying electrons. Thus, in such a material, the transverse resistivity $\rho_{xy}$ is written as the sum of ordinary Hall resistivity $\rho_{xy}^{OHE}$ and the anomalous Hall resistivity $\rho_{xy}^s$. \cite{RevModPhys.82.1539}

\begin{equation}
\rho_{xy}(T)=\rho_{xy}^{OHE}+\rho_{xy}^s = R_0H+R_{AHE}M
\end{equation}

where, $\rho_{xy}^{OHE}$ is written as the product of ordinary Hall coefficient $\mathrm{R_0}$ and the applied magnetic field and $\rho_{xy}^s$ is written as the product of anomalous Hall coefficient $\mathrm{R_{AHE}}$ and magnetization. The ordinary Hall coefficient $\mathrm{R_0}$ depends on the type of carriers (electrons or holes) and their density and is inversely related to the product of the carrier concentration and the electron charge. In lower fields, the measured Hall resistivity is dominated by the AHE, while the effect of OHE typically appears in the higher fields.  

The Hall resistivity($\rho_{xy}$) curves as a function of magnetic field were recorded at various temperatures in the field range of 0 to 50 kOe. Figure \ref{hall_temp} shows the $\rho_{xy}$ versus H curves at 5K, 20 K, 100 K, 160 K, 200 K, 240 K and 300 K for x = 0.1 and 0.3. In the low field regime, $\rho_{xy}$ is found to increase with an increase in magnetic field (dominated by AHE) and gets saturated in the higher field regime. The Hall resistivity shows similar behavior as observed for the magnetization isotherms (see Fig. \ref{MH}). The AHE contribution in $\rho_{xy}$ (i.e. $\rho_{AHE}$) is calculated by extrapolating the high field data towards the zero field since the AHE contribution is typically high and saturates at high fields. Figure \ref{fig:sigma_all} shows the Hall conductivity ($\sigma_{xy}$) versus H curves for x = 0.1 and 0.3. The anomalous Hall conductivity value, $\sigma_{xy0}$ at 5 K is found to be 38 S $\mathrm{cm^{-1}}$ and 43 S $\mathrm{cm^{-1}}$ for x = 0.1 and 0.3 respectively. These values are close to that of the other reported SGS Heusler systems \cite{PhysRevLett.110.100401} and much smaller than those of half-metallic Heusler systems. \cite{PhysRevLett.110.066601}  Figure \ref{fig:CFCG1_hall}(a) and Figure \ref{fig:CFCG3_hall}(a) show the temperature dependence of extracted $\rho_{AHE}$ for x = 0.1 and 0.3 respectively and it is seen that it increases as the temperature is reduced. This behavior is similar to that observed for longitudinal resistivity $\rho_{xx}$(T) (see Fig. \ref{Res}).

The anomalous Hall effect generally has two components: intrinsic and extrinsic.\cite{RevModPhys.82.1539} The intrinsic component is caused by the transverse velocity of Bloch electrons in ideal magnetic crystal and depends only on the band structure. The mechanism of intrinsic contribution to AHE was first proposed by Karplus and Luttinger \cite{PhysRev.95.1154} and later it was redeveloped in terms of Berry's phase. \cite{PhysRevB.59.14915} The extrinsic mechanism is due to the SOC (spin-orbit coupling) induced asymmetric scattering of electrons near impurity sites which give rise to two contributions (i) Skew scattering \cite{SMIT195839}and (ii) side-jump scattering. \cite{PhysRevB.2.4559} To estimate the intrinsic and extrinsic AHE contributions, a scaling model reported by Tian \textit{et al.} \cite{PhysRevLett.103.087206} is used, according to which:
 
\begin{equation}
\rho_{AHE}=\alpha \rho_{xx0}+\beta \rho_{xx0}^2 +b\rho_{xx}^2
\end{equation}

Combining the temperature independent terms, the above equation can be written as:

\begin{equation}
\rho_{AHE}=a'\rho_{xx0}+b\rho_{xx}^2
\end{equation}

where, $a'$= $\alpha +\beta \rho_{xx0}$ and represents the extrinsic contribution due to skew and side jump impurity scattering whereas, b is the intrinsic parameter.
The variation of $\rho_{xy}$ with $\rho_{xx}^2$ along with the fit to equation (9)(blue soild curve) for x = 0.1 and 0.3 is shown in Fig.\ref{fig:CFCG1_hall}(b) and Fig. \ref{fig:CFCG3_hall}(b) respectively. The fitted value of $a'$ is $10^{-2}$ and -1.5 x $10^{-2}$ for x = 0.1 and 0.3 respectively. The value of b which gives the temperature independent intrinsic contribution to $\rho_{AHE}$ is found to increase from 69 S $\mathrm{cm^{-1}}$ to 94 S $\mathrm{cm^{-1}}$ as x changes from x = 0.1 to 0.3. 

Using the relation, $\sigma_{xy}\approx\rho_{xy}/\rho_{xx}^2$, we can calculate the anomalous Hall conductivity $\sigma_{AH}$ and also equation (9) in terms of conductivity can be written as:

\begin{equation}
\sigma_{AHE} = \sigma_{AHE}^{int} + \sigma_{AHE}^{ext}=b+a'\sigma^{-1}_{xx0}\sigma_{xx}^2
\end{equation}

where, $\sigma_{AHE}^{int}$ and $\sigma_{AHE}^{ext}$ represent the intrinsic and extrinsic contributions to the total anomalous Hall conductivity $\sigma_{AHE}$. Here, $\sigma_{xx0} =1/\rho_{xx0}$ and $\sigma_{xx}^2=1/\sigma_{xx}^2$ are the residual and longitudinal conductivity respectively. 
It is important to note that the first term $\sigma_{AHE}^{int}$ is the Karplus-Luttinger term which originates from the Berry curvature and the second term $\sigma_{AHE}^{ext}$ is the sum of extrinsic effects i.e. $\sigma_{AHE}^{ext} = \sigma^{ss}+\sigma^{sj}$ where $\sigma^{ss}$ and $\sigma^{sj}$ represent the extrinsic skew scattering and side jump scattering respectively.\cite{PhysRevLett.103.087206} Figure \ref{fig:CFCG1_hall}(c) and Figure \ref{fig:CFCG3_hall}(c) show the temperature dependence of $|\sigma_{AHE}^{ext}|$, where $\sigma_{AHE}^{ext}$ includes both extrinsic effects (skew scattering and side jump scattering).
The finite $\sigma_{AHE}^{int}$ confirms the non zero Berry curvature in these alloys. The increase in the value of intrinsic parameter \textit{b} with \textit{x} implies an increase in the scattering independent contribution and is attributed to the improved chemical ordering within the lattice.\cite{PhysRevB.89.220406} The negative value of extrinsic parameter $a'$ indicates that the extrinsic contribution due to skew symmetric and the side jump scattering contributed to the $\sigma_{AHE}$ in the opposite way as compared the intrinsic Karplus-Luttinger term does.

\subsubsection{Seebeck coefficient and heat capacity measurements}
Figure \ref{SC} shows the temperature dependence of Seebeck coefficient for $\mathrm{Co_{1.1}Fe_{0.9}CrGa}$ and $\mathrm{Co_{1.3}Fe_{0.7}CrGa}$ alloys. The alloys exhibit positive Seebeck coefficient which indicates that the majority carriers are of hole-type (also confirmed from Hall measurements). The Seebeck coefficient for x = 0.1 (0.3) is negligibly small (nearly zero) in the temperature range from 2 to 40 K (2 to 25 K), which may be attributed to the electron and hole compensation and for $40 < T < 300$ K, it has a very low value of 1.1 $\mu V/K$ (2.2 $\mu V/K$). This behavior is different from that of regular semiconductors, since they are known to have high Seebeck coefficient values ($\sim$ 200 - 300 $\mu$V/K). A similar dependence as seen in the present case was also observed for $\mathrm{Mn_2CoAl}$, another SGS material, from the Heusler family. \cite{PhysRevLett.110.100401} This feature again supports the SGS behavior in these alloys.

Heat capacity measurements were also performed for two alloys x = 0.1 and x = 0.3. In general, the heat capacity can be expressed as the sum of electronic, lattice and magnetic contributions. But at low temperatures, the magnetic excitations have insignificant contribution to the total heat capacity and the other two contributions becomes dominant. Thus, at low temperatures, the heat capacity of a ferromagnetic material can be described by the Sommerfeld-Debye model \cite{PhysRevB.83.235211}, according to which:
\begin{equation}
	C(T)=\gamma T + \beta T^3
\end{equation}
where, the first and second terms represent the electronic and lattice contributions to the total heat capacity respectively. Here, $\gamma$ is the Sommerfeld coefficient representing the electronic part and $\beta$ is the lattice coefficient. Figure \ref{HC} shows the temperature dependence of heat capacity (C/T vs. $\mathrm{T^2}$) for x = 0.1 (left) and x = 0.3 (right) concentration. The inset shows the C vs. T plot in zero field. Evidently, C/T vs. $\mathrm{T^2}$ shows a linear behavior, and as such the slope and the intercept of the curve correspond to the value of $\gamma$ and $\beta$ of Eq.(11) respectively. In the free electron model, the value of $\gamma$ corresponds to the density of states at the Fermi level according to the relation \cite{met7100414}

\begin{equation}
	N(E_F)=\frac{3\gamma}{\pi^2 N_A {k_B}^2}
\end{equation} 
where, $\mathrm{N_A}$ is the Avogadro number and $\mathrm{k_B}$ is the Boltzmann constant.

From the value of $\mathrm{\beta}$, the Debye temperature can be calculated using the following relation\cite{met7100414}
\begin{equation}
	\theta_D = \frac{234ZR}{\beta}
\end{equation}

where, R is the universal gas constant and Z is the number of atoms per formula unit.\\
A fitting of C/T vs. $\mathrm{T^2}$ curve with equation(11) gives $\gamma$ = 18 $\mathrm{m J mole^{-1} K^{-2}}$ and $\beta$ = 0.1 $\mathrm{m J mole^{-1} K^{-4}}$ for x = 0.1. For x = 0.3, the obtained values of $\gamma$ and $\beta$ are 13 $\mathrm{m J mole^{-1} K^{-2}}$ and 0.12 $\mathrm{m J mole^{-1} K^{-4}}$ respectively. The calculated value of density of states $\mathrm{N(E_F)}$ from the extracted Sommerfeld constant for x = 0.1 and 0.3 is found to be 1.9 and 1.4 states $\mathrm{eV^{-1} f.u.^{-1}}$ respectively. The value of $\theta_D$ is found to be 426 K and 405 K for x = 0.1 and 0.3 respectively. It should be noted that, these estimates for N(E$\mathrm{_F}$) and $\theta_D$ are purely based on the free electron model and they only guide us to facilitate a qualitative trend.\\

\subsection{Theoretical Results}

\begin{table}
	\centering
	\caption{Comparison between Slater-Pauling and calculated values of saturation magnetization ($\mathrm{\mu_B/f.u.}$) for $\mathrm{Co_{1+x}Fe_{1-x}CrGa}$ alloys.}
	\begin{ruledtabular}
		\begin{tabular}{c c c}
			
			x   & $M_S(SP)$ & $M_S(Calc.)$\\
			\hline\\
			0.0 & 2.00 & 2.01\\
			0.125 & 2.125 & 2.14\\
			0.25 & 2.25 & 2.25\\
			0.375 & 2.375 & 2.377\\
			0.5 & 2.50 & 2.51\\
			
		\end{tabular}
	\end{ruledtabular}
	\label{tab_mag_theor} 
\end{table}
\begin{figure*}
	\centering
	\includegraphics[width=0.9\linewidth]{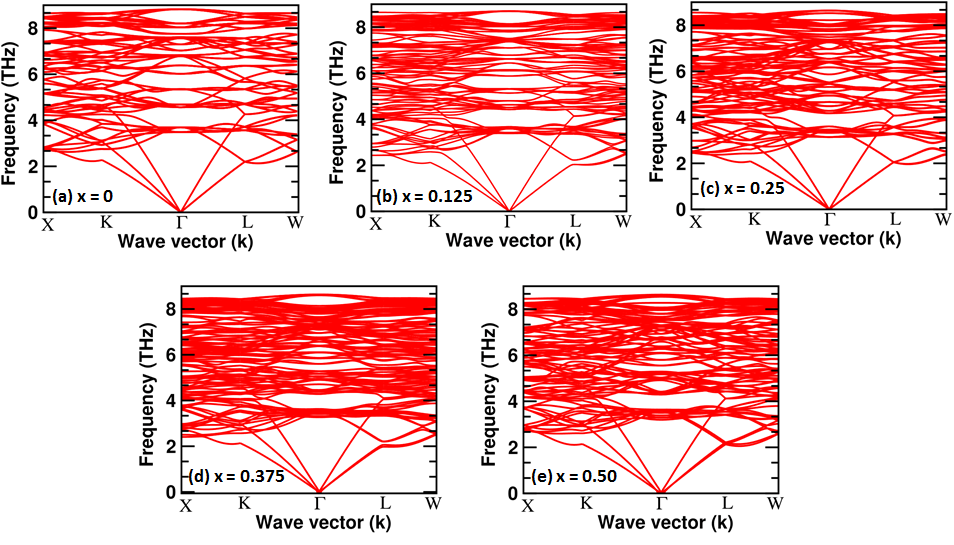}
	\caption{Phonon dispersion of Co$_{1+x}$Fe$_{1-x}$CrGa at different x.}
	\label{phonon}
\end{figure*}

\subsubsection{Electronic structure}

According to the experimental analysis, Co$_{1+x}$Fe$_{1-x}$CrGa (x = 0.1 to 0.5) crystallizes in LiMgPdSn-type structure having space group $F\bar{4}3m$ $(\#216)$. In order to obtain the electronic density of states (DOS) for a desired x, a fully optimized  2x2x2 supercell which corresponds to a minimum ground state energy has been used. 

Figures \ref{dosall} shows the electronic DOS for majority and minority spin channels for Co$_{1+x}$Fe$_{1-x}$CrGa  at x = 0, 0.125, 0.25, 0.375 and 0.5 . At x = 0, i.e. there is no excess of Co, the alloy shows almost vanishing DOS in one spin channel and finite gap in the other spin channel, concluding it to be SGS. The spingapless behavior is retained in the system for $x<0.375$ and the alloy become half metallic for $x\ge0.375$. Although, there are small states at the Fermi level in the spin down channel, but the magnitude of this is negligibly small. The calculated value of total magnetic moment per formula unit matches fairly with the Slater-Pauling value for different x, as shown in Table \ref{tab_mag_theor}.

\subsubsection{Phonon dispersion}	

Phonon dispersion of the alloys, Co$_{1+x}$Fe$_{1-x}$CrGa with various $x$ values, has been shown in Fig.\ref{phonon}. There are 32 atoms in the considered unit cell, which generate three acoustic modes and (3$n$-3 = 93) optical modes, where $n$ is the number of atoms present in the unit cell. We have not observed any imaginary (or soft vibrational modes) frequencies in the whole range of Brillouin zone, which supports thermal stability of the alloys. The non-degeneracy among two transverse acoustic modes and one longitudinal mode can be seen at $K$, $X$ and $W$ high symmetry points. The calculated frequency of the optical modes for Co$_{1+x}$Fe$_{1-x}$CrGa (for all $x$) lie within 3.34 THz to 8.67 THz at $\Gamma$ - point. One interesting features of the phonon dispersion is that some of the optical branches are coupled with the acoustic branches at high symmetry points. This is because of the comparable masses of the constituent elements. The atomic mass of Co, Fe, Cr and Ga is 58.933 amu, 55.845 $amu$, 51.996 $amu$ and 69.723 $amu$ respectively.

\section{Summary and Conclusion}
In conclusion, the quaternary Heusler alloys $\mathrm{Co_{1+x}Fe_{1-x}CrGa}$  are found to show promising properties for spintronic applications. The alloys are found to crystallize in Y-type Heusler structure. The saturation magnetization is found to be in fair agreement with the value predicted by Slater-Pauling rule, which is a prerequisite for spintronic materials. The transition temperature is found to increase with x and lies above room temperature (690 K to 870 K).  Resistivity measurements confirm the semiconducting behavior for x $\leq$ 0.4 and metallic behavior for x = 0.5. However, the absence of exponential dependence of resistivity on temperature indicates the semiconducting nature, but with spin gapless behavior for $x \le 0.4$. The alloy with x = 0.5 shows metallic nature with a minima in resistivity, which is associated with the weak localization effect. ordinary and anomalous Hall contributions have been separated and the intrinsic and extrinsic contribution of the later are also identified. The intrinsic contribution is found to increase as x increases and is correlated with the improved chemical ordering within the lattice. The extrinsic contribution is found to be negative and thus contributes to the AHE in the opposite way as the Karplus-Luttinger term does. The conductivity value ($\mathrm{\sigma_{xx}}$) at 300 K  lies in the range of 2289 S $\mathrm{cm^{-1}}$ to 3294 S $\mathrm{cm^{-1}}$, which is close to other reported SGS materials. The order of magnitude of anomalous Hall conductivity ($\mathrm{\sigma_{AHE}}$) is found to be similar to the other SGS material. The negligible Seebeck coefficient along with the conductivity behavior also supports the SGS nature. Thus, we comclude, on the basis of experiment and theory, that Co$_{1+x}$Fe$_{1-x}$CrGa series show SGS nature up to x = 0.4, beyond which it is HMF. The obtained $\mathrm{T_C}$ value for x = 0.3 (790 K) is found to be the highest among all the previously reported SGS materials.

\section*{Acknowledgments}
Deepika Rani would like to thank Council of Scientific and Industrial Research (CSIR), India for providing Senior Research Fellowship. DR would also like to thank Dr. V. K Kushwaha for helping in Seebeck coefficient measurements.

\bibliography{bib}

\end{document}